\documentclass[%
 reprint,
 amsmath,amssymb,
 aps,
 prl,
]{revtex4-1}

\usepackage[dvips]{graphicx}
\usepackage{dcolumn}
\usepackage{bm}

\usepackage{color,graphicx}
\usepackage[colorlinks=true,urlcolor=blue,citecolor=blue,linkcolor=blue]{hyperref}
\usepackage{textcomp}
\usepackage{pdfpages}

\makeatletter
\AtBeginDocument{\let\LS@rot\@undefined}
\makeatother

\definecolor{ROT}{rgb}{0.7,0.1,0.2}

\begin{document}
\preprint{APS/123-QED}

\title{Anomalous Quantum Correlations of Squeezed Light}

\author{B. K\"uhn}

\author{W. Vogel}

\affiliation{Arbeitsgruppe Theoretische Quantenoptik, Institut f\"ur Physik, Universit\"at Rostock, D-18051 Rostock, Germany}

\author{M. Mraz}

\author{S. K\"ohnke}

\author{B. Hage}

\affiliation{Arbeitsgruppe Experimentelle Quantenoptik, Institut f\"ur Physik, Universit\"at Rostock, D-18051 Rostock, Germany}

\date{\today}

\begin{abstract}


Three different noise moments of field strength, intensity, and their correlations are simultaneously measured.
For this purpose a homodyne cross-correlation measurement~\cite{Vo95}
is implemented by superimposing the signal field and a weak local oscillator on an unbalanced beam splitter.
The relevant information is obtained via the intensity noise correlation of the output modes.
Detection details like quantum efficiencies or uncorrelated dark noise are meaningless for our technique.
Yet unknown insight in the quantumness of a squeezed signal field is retrieved from the anomalous moment, correlating field strength with intensity noise.
A classical inequality including this moment is violated for almost all signal phases. 
Precognition on quantum theory is superfluous, as our analysis is solely based on classical physics. 

\end{abstract}

\pacs{Valid PACS appear here}
\maketitle

\paragraph{Introduction.---}

To distinguish nonclassical effects of light from classical ones and to conceive possible applications has been a central question of quantum optics for several decades. 
It is of fundamental interest if the outcome of an optical experiment can
be interpreted in the framework of classical statistical electrodynamics, or if a quantum description is necessary. 
A possible way to certify nonclassical effects is based on moments, as, e.g., quadrature squeezing~\cite{Slu85,Kim86} or sub-Poisson statistics~\cite{Man83}, each is based on a single observable quantity.

In contrast, anomalous moments composed of noncommuting observables are hard to access in experiments.
An important example is the correlation of intensity and field strength noise, as it unifies the particle and wave nature of quantum light.
Its measurement was originally proposed by a homodyne correlation technique with a weak local oscillator (LO)~\cite{Vo91}.
Anomalous moments were detected in resonance fluorescence of a single trapped atom~\cite{Bla1}. In this setting, balanced homodyne detection (BHD) with a weak LO was conditioned on the detection of a resonance 
fluorescence photon. 
Conditional homodyne detection was also studied by simulations~\cite{Car1} and experiments~\cite{Fos1},
which allows us to observe large violations of a Schwarz inequality; 
see also Ref.~\cite{Fos2}. 
However, this approach only applies to a Gaussian or weak source field and it requires three detectors. 
Higher-order correlations of multiple field modes are accessible by balanced or unbalanced homodyne correlation measurements~\cite{Shc06,Kue16}. 

In Ref.~\cite{Vo95}, two detection schemes have been theoretically analyzed, which use four-port homodyning with comparable intensities of signal and LO.
One of the techniques, called homodyne intensity correlation measurement, was introduced in~\cite{Vo91}. 
It was realized only recently to certify quadrature squeezing in resonance fluorescence light from a single quantum dot~\cite{Schulte15}.
Negative values of the measured intensity noise correlation directly uncover nonclassicality of the signal field.
The other technique in~\cite{Vo95} was called homodyne cross-correlation measurement (HCCM): signal and LO are interfered at a single unbalanced beam splitter and the 
two output fields are recorded with linear detectors. 
Unlike in BHD, a correlation measurement is performed. The
detector currents are multiplied and not subtracted, which yields second-order intensity noise correlations.
An experimental realization of this method has been missing. 

In the present Letter, we report the first experimental implementation of the HCCM.
Our signal field is prepared in a phase-squeezed coherent state, generated via parametric down-conversion. 
For the intensity regime we use for signal and LO standard linear photodiodes are suitable.
The contributions of different orders of the LO field strength are extracted from the measured correlation function. 
Our method certifies anomalous quantum correlations of squeezed light even for most of the antisqueezed phase region.

\begin{figure}[ht]
	\includegraphics*[width=8.00cm]{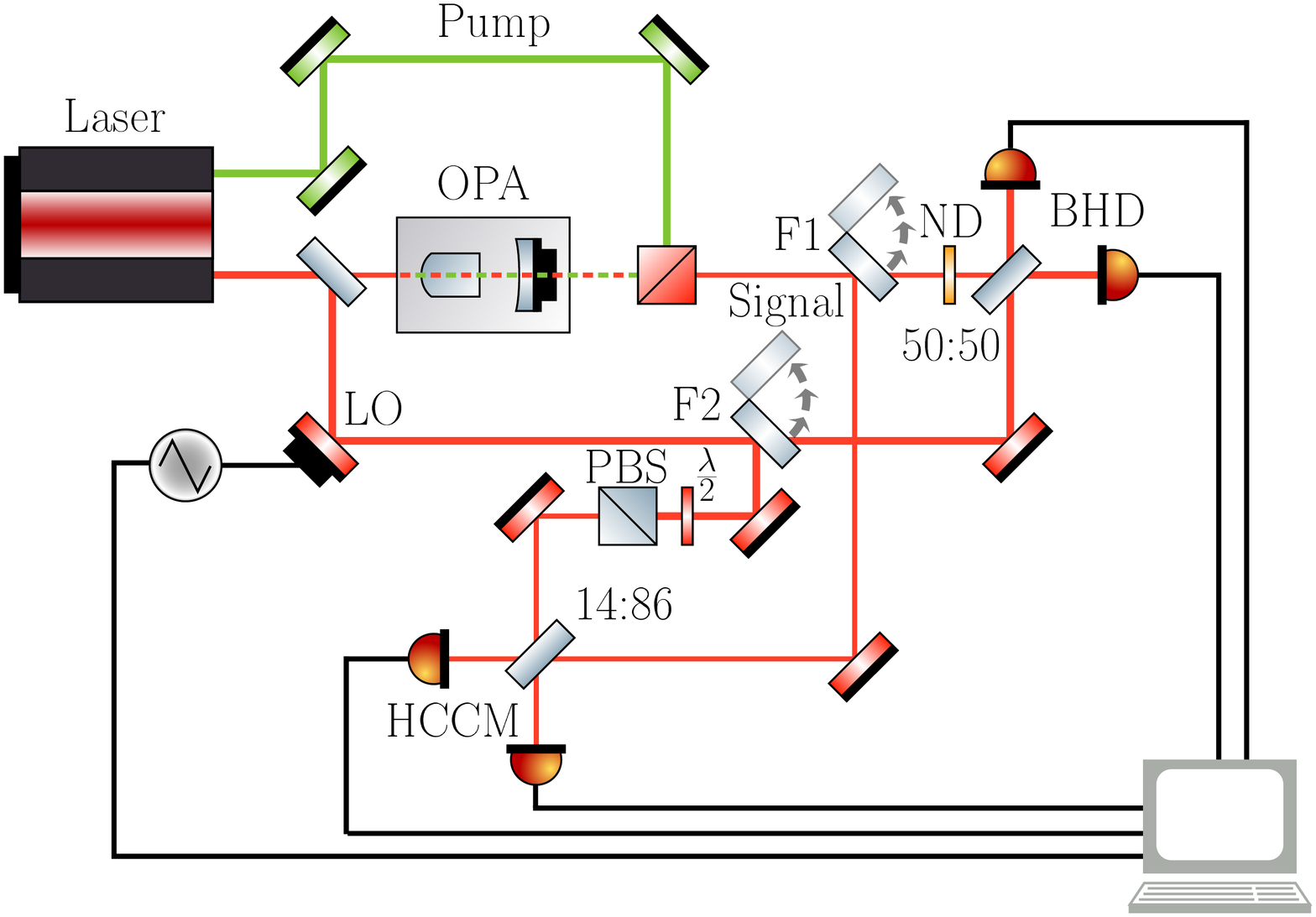}
	\caption{Experimental setup for the generation and detection of squeezed light.
		The squeezed field is generated in an OPA.
		The flip mirrors F1 and F2 are used to send the squeezed field either to the BHD
		or the HCCM device.
	}\label{fig:ExpSetup}
\end{figure}

\paragraph{Homodyne cross-correlation measurement.---}
The basic setup of our measurement technique is illustrated in Fig.~\ref{fig:ExpSetup}.
The investigated squeezed field was generated in a hemilithic, standing wave, nonlinear cavity, used as an optical parametric amplifier (OPA).
An 11\,\rm{mm} long 7\% magnesium oxide-doped lithium niobate (7\%MgO:LiNbO$_{3}$) crystal served as a $\chi^{(2)}$-nonlinear medium with noncritical phase matching.
A strong seed beam was inserted into the OPA to produce a coherently displaced squeezed field with a signal power of 284\,\textmu\rm{W}.
The OPA was pumped with 243\,\rm{mW} at 532\,\rm{nm} resulting in a gain of 2.3 at 1064\,\rm{nm}. 
For the HCCM the LO power is of the magnitude of the signal power. Both fields are combined on an unbalanced beam splitter and the two output beams are recorded with photodetectors (PDs).
For an independent state characterization we used the established method of BHD.
There is only one difference to a normal BHD device, an ND filter is placed in the signal beam in front of the 50:50 beam splitter to reduce the intensity of the signal to 32\,\textmu\rm{W}. This avoids 
demolition of the PDs, as the LO power can be reduced to 1.03\,\rm{mW}. 
Because of the knowledge of the power reduction in the signal field, we are able to estimate the squeezing of the undamped signal to be -2.7\,\rm{dB} and the antisqueezing to be 5.5\,\rm{dB}.
The visibility in the BHD setup is 97\,\% and the quantum efficiency $\sim90$\,\%.
In the HCCM setup the visibility is 96\,\% and the quantum efficiencies of the PDs are $\sim94$\,\%.
In both detection setups we used the technique of continuous variation of the optical phase as presented in~\cite{Agu15}.
This provides a uniformly distributed phase.

The measurement outcome of the HCCM is the correlation of electric current fluctuations (ac) of the two detectors.
The ac time sequences, $[c_1^{(1)} \cdots c_1^{({N_\phi})}]_\phi$ and $[c_2^{(1)} \cdots c_2^{({N_\phi})}]_\phi$, measured
for a particular LO phase $\phi$, are same-time correlated, i.e.,
\begin{eqnarray}\label{eq:measuredcorr}
C(\phi)=\overline{c_{\mathrm{1}}(\phi) c_{\mathrm{2}}(\phi)}=\dfrac1{N_\phi}\sum_{\ell=1}^{N_{\phi}} c^{(\ell)}_{1}(\phi)\,c^{(\ell)}_{2}(\phi). 
\end{eqnarray}
For the intensities present in our experiment, the detectors respond linearly.
Therefore, the quantity~\eqref{eq:measuredcorr} is proportional to the intensity noise correlation $\Delta G^{(2,2)}(\phi)=\langle\Delta I_1\Delta I_2\rangle_\phi$, i.e., 
\begin{eqnarray}\label{eq:CG}
C(\phi)=\zeta_1\zeta_2\Delta G^{(2,2)}(\phi),
\end{eqnarray}
where $\langle\cdot\rangle$ is the classical expectation value and $\zeta_k$ is the product of detector parameters such as detector efficiency, gain factor, 
and other positive scaling factors of the detectors $k=1,2$.
The intensity noise correlation can be separated into three contributions with different powers of the LO field strength $E_{L}$,
\begin{eqnarray}\label{eq:fundamental}
\Delta G^{(2,2)}(\phi)=\Delta G_0^{(2,2)}+\Delta G_1^{(2,2)}(\phi)+\Delta G_2^{(2,2)}(\phi).
\end{eqnarray}
Defining coefficients $\mathcal{T}_i$ by $(\mathcal{T}_0,\mathcal{T}_1,\mathcal{T}_2)=(1,|R|/|T|-|T|/|R|,-1)$ and $\mathcal{T}=|T|^2|R|^2$,
the zeroth-order (in $E_{L})$ term is given by
\begin{eqnarray}\label{eq:contribution0}
\Delta G_0^{(2,2)}=\mathcal{T}\mathcal{T}_0\langle(\Delta I)^2\rangle
\end{eqnarray}
with the signal intensity $I=E^{(-)}_\phi E^{(+)}_\phi$ and the intensity noise $\Delta I=I-\langle I\rangle$. It is independent of both phase and field strength of the LO.
The first-order term, 
\begin{eqnarray}\label{eq:contribution1}
\Delta G_1^{(2,2)}(\phi)=\mathcal{T}\mathcal{T}_1E_{L}\langle\Delta E_\phi\Delta I\rangle, 
\end{eqnarray}
with the signal (electric) field strength $E_\phi=E^{(+)}_\phi+E^{(-)}_\phi$ and the corresponding fluctuation
$\Delta E_\phi=E_\phi - \langle E_\phi\rangle $, in general is $2\pi$ periodic in the phase and linear in the field strength of the LO.
Note that this anomalous moment is composed of two 
observables.
A Fourier decomposition of the second-order term,
\begin{eqnarray}\label{eq:contribution2}
\Delta G_2^{(2,2)}(\phi)=\mathcal{T}\mathcal{T}_2E_{L}^2\langle(\Delta E_\phi)^2\rangle,
\end{eqnarray}
which is quadratic in the LO field strength, is in general composed of a $\pi$-periodic and a constant Fourier component in the LO phase. 
The different dependences of the terms~\eqref{eq:contribution0}--\eqref{eq:contribution2} on the phase and field strength of the LO allow
us to separate them from $\Delta G^{(2,2)}(\phi)$; for details see~\cite{Vo95} and the discussion below.

Additional contributions in~\eqref{eq:fundamental} arise from classical fluctuations of the LO, which though very small in our case are evaluated as follows.
The dominant effect is a constant offset, obtained from a correlation measurement with blocked signal. 
This yields a direct observation of the intensity fluctuation of the LO, including possibly occurring correlated dark noise in the two detectors. 
To correct for LO and correlated dark noise, this offset is removed from the correlation $C(\phi)$ measured in the case with unblocked signal.
A strong point of the technique is that even if uncorrelated dark noise in both detectors were stronger than the quantum noise of the signal, it does not contaminate the measurement result.
By contrast, uncorrelated dark noise is relevant in BHD.

Note that the expressions~\eqref{eq:contribution0}--\eqref{eq:contribution2} are also correct for a lossy beam splitter, i.e., $|T|^2+|R|^2<1$.
The theory of Ref.~\cite{Vo95} can also be extended to an asymmetric beam splitter; see, e.g.,~\cite{Upp16}.
In this case, the intensity reflection-transmission ratio of the beam splitter for the LO ($|R_L|^2:|T_L|^2$) and 
for the signal ($|R_S|^2:|T_S|^2$) can be different. 
This yields the more general coefficients $(\mathcal{T}_0,\mathcal{T}_1,\mathcal{T}_2)=((|R_S|/|R_L|)(|T_S|/|T_L|),|R_S|/|T_L|-|T_L|/|R_S|,-1)$ 
and $\mathcal{T}=|T_S||T_L||R_S||R_L|$. Our beam splitter shows symmetric transmittance, i.e., $|T_S|^2=|T_L|^2$, but asymmetric reflectance, i.e., $|R_S|^2\neq|R_L|^2$.

If the LO is strong compared with the signal, the term $\Delta G_2^{(2,2)}(\phi)$ is dominant, and the correlation outcome is proportional to the negative squeezing effect.
Accordingly, in this scenario the anomalous moment negligibly contributes to the total correlation and it is, therefore, not accessible.
Even if the LO intensity is comparable to the signal intensity, the anomalous moment is only accessible if the beam splitter is unbalanced~\cite{Vo95}. 
The maximum visibility is reached for a  $14:86$ intensity partition, which is approximately used in our experiment.

\paragraph{Separation of moments.---}
Let us study the separation of the contributions 
\begin{eqnarray}
C_k(\phi)=\zeta_1\zeta_2\Delta G^{(2,2)}_k(\phi)
\end{eqnarray}
from the total correlation $C(\phi)$, which is given by a second degree trigonometric polynomial,
\begin{eqnarray}\label{eq:fit}
C(\phi)=a_0+\sum_{k=1}^2\left[a_k\cos(k\phi)+b_k\sin(k\phi)\right], 
\end{eqnarray}
with real parameters $a_k$ and $b_k$, as proposed in~\cite{Vo95}.
Since both $C_0(\phi)$ and $C_2(\phi)$ contain a phase-independent part,
it is necessary to perform in addition a measurement with blocked LO, which yields the resulting correlation outcome $C_{\mathrm{block}}$.
The contributions $C_k(\phi)$ are obtained from the latter and the Fourier coefficients as
\begin{eqnarray}
C_0(\phi)&=&C_{\mathrm{block}}\label{eq:C0}\\
C_1(\phi)&=&a_1\cos(\phi)+b_1\sin(\phi)\label{eq:C1}\\
C_2(\phi)&=&a_2\cos(2\phi)+b_2\sin(2\phi)+a_0-C_{\mathrm{block}}\label{eq:C2}. 
\end{eqnarray}

\begin{figure}[ht]
        \hspace{-0.5cm}
	\includegraphics*[width=9.0cm]{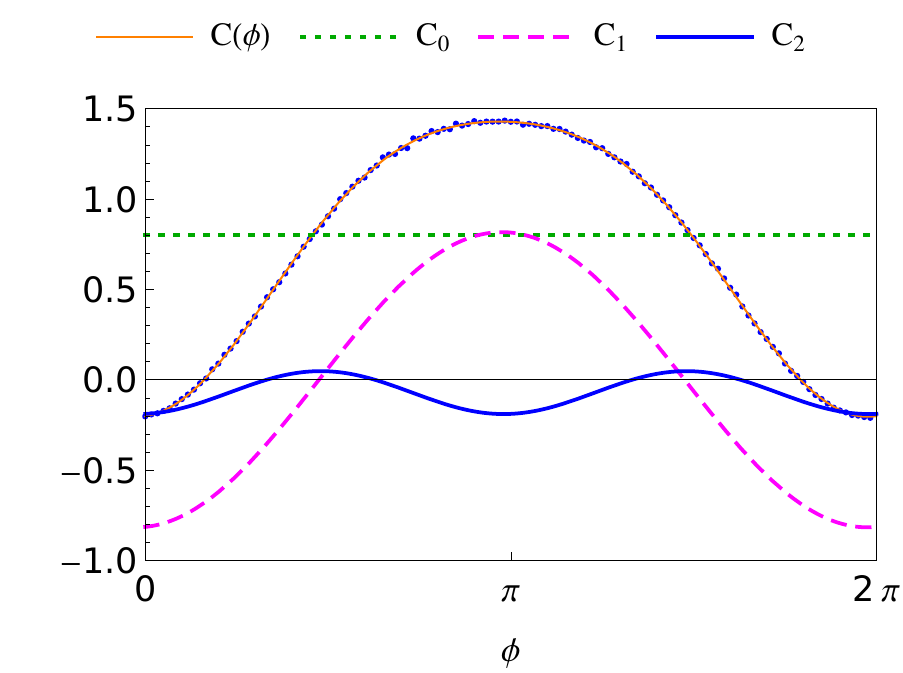}
 	\caption{Measured correlation $C(\phi)$ (markers) as a function of phase.
	The error bars of one standard deviation statistical uncertainty are within the size of the markers.
	The fit according to Eq.~\eqref{eq:fit} is shown by the thin solid curve,
	composed of the contributions $C_0$, $C_1$, and $C_2$.} 
	\label{fig:contributions}
\end{figure}

Figure~\ref{fig:contributions} shows the measured correlation $C(\phi)$ for 
$120$ phases selected equidistantly in $[0,2\pi]$ and the fit according to Eq.~\eqref{eq:fit}. 
For each phase the same number of $4.58\times 10^{5}$ data samples was used.
For details on the fit via regression analysis~\cite{Raw98,Wei05} and the error calculation see Supplemental Material.
We observe an excellent agreement of the experimental outcome with the theoretical prediction.
For the LO-blocked case we obtain $C_{\mathrm{block}}=0.80153\pm0.00014$ using $3\times 10^8$ data samples.
In addition, the extracted contributions $C_0$, $C_1$, and $C_2$ are shown. 
One clearly observes the $2\pi$-periodic anomalous moment of intensity-field noise.
Once a calibration of the setup is performed, i.e., $\zeta_1$ and $\zeta_2$ in Eq.~\eqref{eq:CG} and the LO strength are known, the moments can be quantified.

It is important to note that our method is quite sensitive to drifts of the signal state, since one has to ensure that approximately the same signal state is present in the LO-blocked and unblocked case.
We incorporate a drift error of $C_{\mathrm{block}}$ as the difference of the result of two subsequent measurements.
Note that drift errors can be further reduced by increasing the frequency of blocking and unblocking the LO. 

\begin{figure}[ht]
        \hspace{-0.5cm}
	\includegraphics*[width=9.0cm]{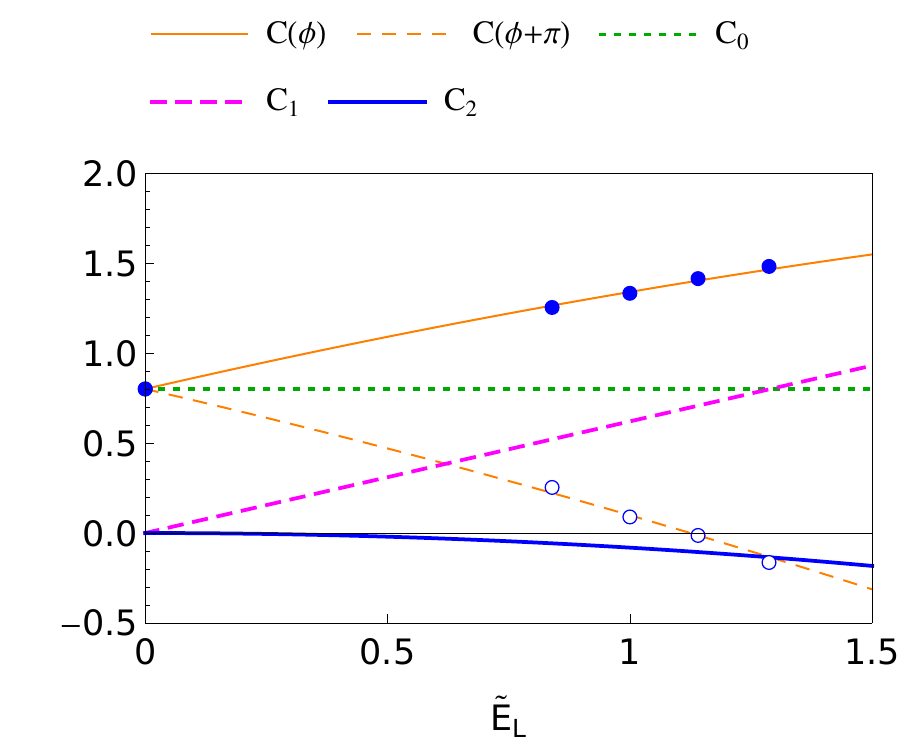}
	\caption{Measured correlation $C(\phi)$ (filled markers) and $C(\phi+\pi)$ (unfilled markers) for $\phi=3\pi/4$ 
	as a function of the rescaled LO field strength 	$\tilde E_L$.
	The error bars of one standard deviation statistical uncertainty are within the size of the markers.
	The quadratic fits are the thin solid and dashed curves,
	composed of the contributions $C_0$, $C_1$, and $C_2$.} 
	\label{fig:LO}
\end{figure}

Alternatively, the contributions $C_k$ may be separated by the dependence on the LO field strength; see Supplemental Material.
In our experiment five different LO powers namely 0 (blocked LO), 117, 166, 216, and 275\,\textmu\rm{W} were probed for the phases $\phi=3\pi/4$ and $\phi+\pi$.
The result is shown in Fig.~\ref{fig:LO} together with the contributions $C_k$ proportional to $E_L^k$.

\paragraph{Classical correlations.---}
In a classical picture an inequality can be derived based on the extracted moments, which is always fulfilled.
For an arbitrary function $f$ of $E_\phi^{(\pm)}$, the expectation value $\langle |f|^2\rangle$ is non-negative.
For our experimental outcome we use a properly chosen function of the form
$f=h_0\Delta I+h_1 
\Delta E_\phi$ and $h_0,h_1\in\mathbb{C}$. 
Defining the matrix
\begin{eqnarray}
M(\phi)=
\begin{pmatrix}
\langle(\Delta  I)^2\rangle & 
\langle\Delta E_\phi\Delta  I\rangle \\
\langle\Delta E_\phi\Delta  I\rangle & 
\langle(\Delta  E_\phi)^2\rangle 
\end{pmatrix},
\end{eqnarray}
the determinant of $M(\phi)$ for a classically correlated signal field is non-negative for all phases $\phi\in[0,2\pi)$. This is equivalent to the inequality
\begin{eqnarray}\label{eq:det}
\langle\Delta E_\phi\Delta I\rangle^2 \leq \langle(\Delta I)^2\rangle\langle(\Delta E_\phi)^2\rangle.
\end{eqnarray}
If the beam splitter transmittance and reflectance ratios are known, one can determine the matrix
\begin{eqnarray}\label{eq:L}
L(\phi)&=&
\begin{pmatrix}
C_0(\phi)/\mathcal{T}_0 & C_1(\phi)/\mathcal{T}_1 \\
C_1(\phi)/\mathcal{T}_1 & C_2(\phi)/\mathcal{T}_2
\end{pmatrix}
\end{eqnarray}
from the contributions $C_k(\phi)$
of $C(\phi)$. The determinant of this matrix is related to the determinant of $M(\phi)$ as
\begin{eqnarray}\label{eq:dets}
\mathrm{det}\left[L(\phi)\right]=\zeta_1^2\zeta_2^2\mathcal{T}^2E_{L}^2\,\mathrm{det}\left[M(\phi)\right]. 
\end{eqnarray}

Obviously, the sign of $\mathrm{det}\left[L(\phi)\right]$ equals that of $\mathrm{det}\left[M(\phi)\right]$. Thus, 
the necessary condition~\eqref{eq:det} for a classically correlated signal field can be tested directly by the matrix $L(\phi)$ 
through $\mathrm{det}[L(\phi)]\geq 0$. Note that 
no knowledge of the efficiencies and gain factors incorporated in the detection process is required. 
Also the exact strength of the (weak) LO is meaningless, cf.,~Eq.~\eqref{eq:dets}. 
The ratios $|R_L|^2:|T_L|^2$ and $|R_S|^2:|T_S|^2$ have to be known, but not the reflectance and transmittance itself, which makes the test robust to beam splitter losses.

\begin{figure}[ht]
\centering
 	\includegraphics*[width=8.30cm]{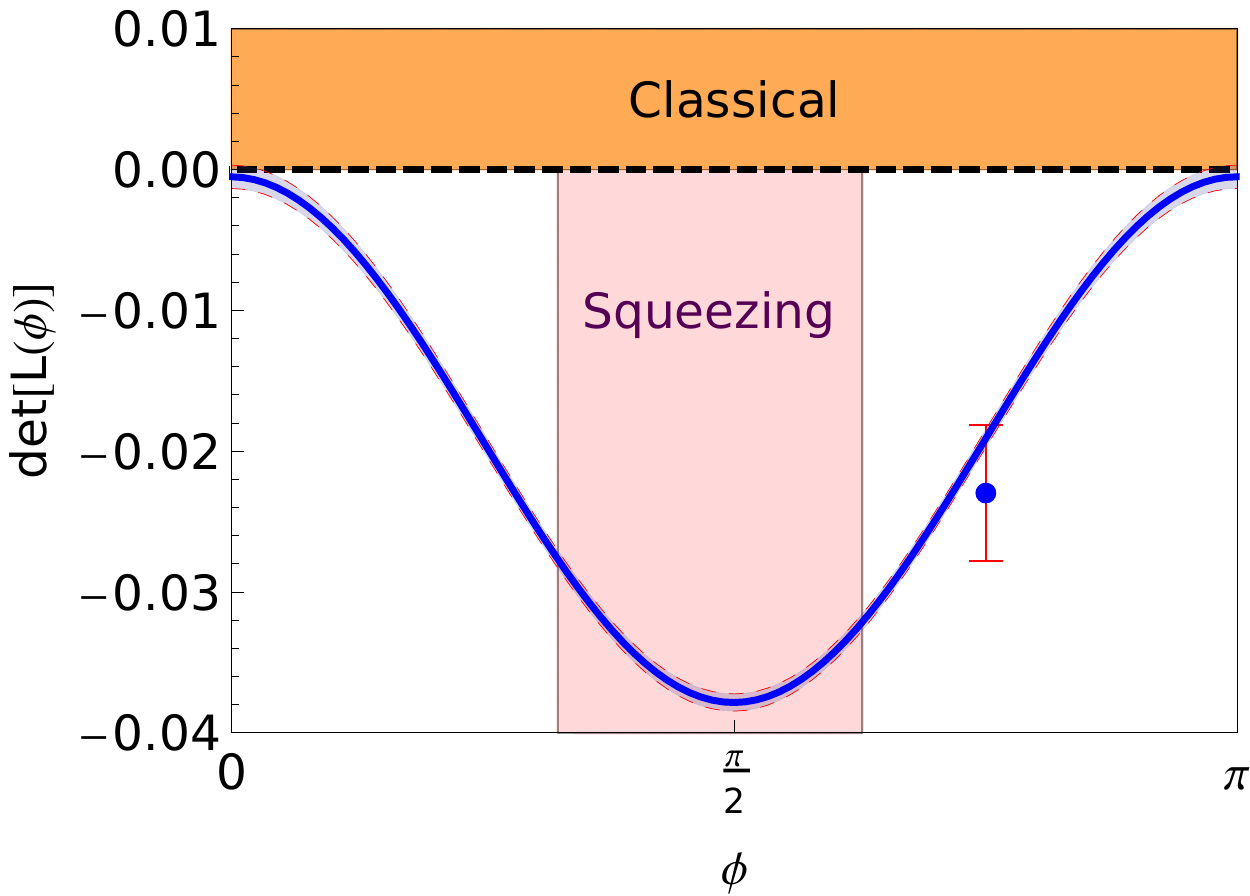}
	\caption{The solid line shows $\mathrm{det}[L(\phi)]$ as a function of phase as obtained through separation by different phase periodicity.
	Because of the $\pi$ periodicity of the plot, we confine ourselves to the interval $[0,\pi]$.
	The thin dashed lines correspond to an error of one standard deviation. The thick dashed line marks the border between the classical and nonclassical regions. 
	Squeezing is present within the light-colored interval. 
	The marker at $\phi=3\pi/4$ (antisqueezed region) follows from the separation by the LO field strength, from   data measured for $\phi$ and $\phi+\pi$. The error corresponds to one standard deviation.}
	\label{fig:nclresult}
\end{figure}

\paragraph{Quantum correlations.---}
Figure~\ref{fig:nclresult} shows the experimental result for $\mathrm{det}[L(\phi)]$ as a function of the LO phase.
The determinant is significantly negative in a wide range of phases $\phi$, which is a clear violation of the classicality condition~\eqref{eq:det}. 
Remarkably, the determinant is even negative for phases where no squeezing is present, e.g., for $\phi=3\pi/4$ with $28$ standard deviations significance. 
Hence the anomalous quantum correlations under study also exist in the antisqueezed phase region.
For comparison, the determinant obtained by separation through the LO field strength dependence is shown for $\phi=3\pi/4$.
Since the LO intensity is not scanned continuously in our case, the drift of the signal state yields a larger uncertainty than the separation by phase. 
Nevertheless, this proof-of-principle experiment certifies nonclassicality with a significance of $4.7$ standard deviations.
With some technical effort, this technique could also be further improved. 

Our method is especially beneficial, when the phase interval of squeezing is small, e.g., for strong squeezing or phase diffused states.
Then it is challenging to stabilize the system onto the squeezed phase. In this regard, our method may detect quantum effects under demanding squeezing conditions of the input state. 
Note that the positive correlation outcome $C_{\mathrm{block}}$ for blocked LO shows that
the necessary classicality condition $\langle(\Delta I)^2\rangle\geq 0$ for the variance of the signal intensity is valid.


It is important that the whole previous analysis is purely classical and does not require any bosonic commutation relations~\cite{Aga68,Cah69,AgaI70,AgaII70}.
This essential property has the benefit that 
the derived classicality condition based on anomalous correlations applies without 
assumptions on the validity of quantum physics for the interpretation of the measurement outcome. 
By contrast, the squeezing condition $\langle(\Delta \hat E_\phi)^2\rangle<\langle(\Delta\hat E_\phi)^2\rangle_{\mathrm{vac}}$ for a particular phase $\phi$, 
which is applied in balanced homodyne detection,
intrinsically utilizes nonvanishing commutators. Hence, such quantumness tests require the postulate of the validity of quantum physics.
This consideration is closely related to the definition of nonclassicality in the sense of Titulaer and Glauber~\cite{Tit1}, which is based on the Glauber-Sudarshan $P$~function~\cite{Gla63,Sud63}.
That is, a state is nonclassical if it violates a condition $\langle:\hat f^\dagger\hat f:\rangle\geq 0$,
wherein $:\cdot:$ denotes normal ordering, classical expectation values are replaced by the quantum mechanical ones, and classical field quantities $f$ are replaced by the corresponding field operators $\hat f$.

It is eminent that our HCCM device accesses, based on quantum measurement theory~\cite{Vo91,Vo95}, three pairwise noncommuting observables, 
$:(\Delta\hat I)^2:$, $:(\Delta\hat E_\phi)^2:$, and $:\Delta\hat E_\phi\hat I:$, within a single measurement scenario. 
The anomalous correlations violating the classicality condition~(\ref{eq:det}), cf.,~Fig.~\ref{fig:nclresult}, turn out to be in excellent agreement with the condition for anomalous quantum correlations,
\begin{eqnarray}\label{eq:an-qu-corr}
\langle:\Delta\hat E_\phi\Delta \hat I:\rangle^2 > \langle:(\Delta \hat I)^2:\rangle\langle:(\Delta \hat E_\phi)^2:\rangle,
\end{eqnarray}
of the normal-ordered fluctuations of intensity and field strength. For the derivation of general criteria for quantum correlations of light, we refer to~\cite{Vo08}.

\paragraph{Conclusions.---}
In conclusion, we have experimentally realized the homodyne cross-correlation measurement to observe up to fourth-order moments of the field fluctuations of a phase-squeezed coherent state.
In particular, this allows us to determine the anomalous moment, which is composed of two noncommuting observables, namely, intensity and field strength noise, which is observed with high significance.
Furthermore, a quantum correlation test based on solely the measured moments shows the existence of anomalous quantum correlations even outside the squeezed phase region.
As a central benefit, the data analysis of our technique is completely free of quantum physical assumptions, such as nonvanishing commutation relations. Hence the technique visualizes directly violations 
of classical physics. The anomalous quantum correlations of squeezed light, which have been verified here for the first time, 
may pave the way for alternative applications of squeezed light in quantum technology, beyond the phase interval of squeezing.

\begin{acknowledgements}
The authors are grateful to Oskar Schlettwein for valuable discussions.
This work has been supported by the European Commission through the project QCUMbER (Grant No. 665148) and by the Deutsche Forschungsgemeinschaft through SFB 652 (Grants No. B12 and No. B13). 
\end{acknowledgements}

\begin{widetext}
\newpage


\section*{Supplementary material}

# Supplemental Material
# Anomalous Quantum Correlations of Squeezed Light

B. Kühn and W. Vogel

*Arbeitsgruppe Theoretische Quantenoptik, Institut für Physik, Universität Rostock, D-18051 Rostock, Germany*

M. Mraz, S. Köhnke, and B. Hage

*Arbeitsgruppe Experimentelle Quantenoptik, Institut für Physik, Universität Rostock, D-18051 Rostock, Germany*

(Dated: April 25, 2017)

Section I outlines the separation of the moments from the correlation of the two measured detector-current fluctuations by using the different LO phase dependence. The technique to separate these moments by the LO field strength is considered in Section II.

## I. SEPARATION OF MOMENTS BY LO PHASE

In this section the separation of moments using the different dependence on the LO phase is studied. Consider $K$ points $(\phi_j, C(\phi_j))$, $j = 1, \ldots, K$ for different phases $\phi_j$, which are obtained independently from $N_{\phi_j}$ data points according to Eq. (1) of the Letter. The standard error of the mean of $C(\phi_j)$ is

$$u_{C(\phi_j)} = \sqrt{\sum_{\ell=1}^{N_{\phi_j}} \frac{[c_1^{(\ell)}(\phi_j)c_2^{(\ell)}(\phi_j) - C(\phi_j)]^2}{N_{\phi_j}(N_{\phi_j} - 1)}}. \tag{1}$$

The statistical uncertainty of $C_{\text{block}}$ is obtained analogously by

$$u_{C_{\text{block}}} = \sqrt{\sum_{\ell=1}^{N_{\text{block}}} \frac{[c_{1,\text{block}}^{(\ell)} c_{2,\text{block}}^{(\ell)} - C_{\text{block}}]^2}{N_{\text{block}}(N_{\text{block}} - 1)}}, \tag{2}$$

where $N_{\text{block}}$ is the number of data samples recorded for the LO-blocked scenario. This uncertainty is increased by the error arising from a drift of the signal state. The measured values are fitted by

$$C(\phi) = a_0 + \sum_{k=1}^{2} [a_k \cos(k\phi) + b_k \sin(k\phi)] \tag{3}$$

which corresponds to the minimization of the Euclidean norm $\|U\boldsymbol{a} - \boldsymbol{C}\|_2$ with respect to the vector of real fit parameters $\boldsymbol{a} = (a_0, a_1, b_1, a_2, b_2)^T$. Here

$$U = \begin{pmatrix} 1 & \cos(\phi_1) & \sin(\phi_1) & \cos(2\phi_1) & \sin(2\phi_1) \\ 1 & \cos(\phi_2) & \sin(\phi_2) & \cos(2\phi_2) & \sin(2\phi_2) \\ \vdots & \vdots & \vdots & \vdots & \vdots \\ 1 & \cos(\phi_K) & \sin(\phi_K) & \cos(2\phi_K) & \sin(2\phi_K) \end{pmatrix} \tag{4}$$

is a $K \times 5$ dimensional matrix and $\boldsymbol{C} = (C(\phi_1), \cdots, C(\phi_K))^T$. The solution of the problem is given by

$$\boldsymbol{a} = (U^T U)^{-1} U^T \boldsymbol{C} \tag{5}$$

(see e.g. [1, 2]) and the statistical uncertainty

$$(u_{\boldsymbol{a}})_i = \sqrt{\sum_{j=1}^{K} [((U^T U)^{-1} U^T)_{ij} (u_{\boldsymbol{C}})_j]^2} \tag{6}$$

with $i = 1, \ldots, 5$ follows from error propagation. Further error analysis yields for the components $C_k(\phi)$ (cf. Eqs. (9)–(11) of the Letter) of $C(\phi)$ the uncertainties

$$u_{C_0(\phi)} = u_{C_{\text{block}}} \tag{7}$$

$$u_{C_1(\phi)} = \sqrt{u_{a_1}^2 \cos^2(\phi) + u_{b_1}^2 \sin^2(\phi)} \tag{8}$$

$$u_{C_2(\phi)} = \sqrt{u_{a_2}^2 \cos^2(2\phi) + u_{b_2}^2 \sin^2(2\phi) + u_{a_0}^2 + u_{C_{\text{block}}}^2}. \tag{9}$$

The error of the quantities

$$B_k(\phi) = \frac{C_k(\phi)}{\mathcal{T}_k}, \tag{10}$$

which are required for the matrix $L(\phi)$ (cf. Eq. (14) of the Letter), is given by

$$u_{B_k(\phi)} = \frac{1}{|\mathcal{T}_k|}\sqrt{[B_k(\phi)]^2 u_{\mathcal{T}_k}^2 + u_{C_k(\phi)}^2} \tag{11}$$

with

$$u_{\mathcal{T}_0} = \sqrt{t^2 u_r^2 + r^2 u_t^2} \tag{12}$$

$$u_{\mathcal{T}_1} = \left(1 + \frac{1}{m^2}\right) u_m \tag{13}$$

$$u_{\mathcal{T}_2} = 0. \tag{14}$$

Here we define the ratios

$$r = \frac{|R_S|}{|R_L|} = \frac{s}{\ell} \tag{15}$$

$$t = \frac{|T_S|}{|T_L|} = 1 \tag{16}$$

$$m = \frac{|R_S|}{|T_L|} = s \cdot t \tag{17}$$

with the uncertainties

$$u_r = \frac{1}{\ell}\sqrt{u_s^2 + r^2 u_\ell^2} \tag{18}$$

$$u_t = 0 \tag{19}$$

$$u_m = \sqrt{t^2 u_s^2 + s^2 u_t^2}, \tag{20}$$

as well as the ratios

$$s = \frac{|R_S|}{|T_S|} \tag{21}$$

$$\ell = \frac{|R_L|}{|T_L|}, \tag{22}$$

together with their uncertainties $u_s$ and $u_\ell$, which are determined by measuring the DC-levels of the two detectors and incorporating a correction in case of different quantum efficiencies. Finally, one obtains for the statistical error of $\det[L(\phi)]$

$$u_{\det[L(\phi)]} = \sqrt{[B_2(\phi)]^2 u_{B_0(\phi)}^2 + [B_0(\phi)]^2 u_{B_2(\phi)}^2 + 4[B_1(\phi)]^2 u_{B_1(\phi)}^2}. \tag{23}$$

The signed statistical significance is given by

$$\Sigma = \frac{\det[L(\phi)]}{u_{\det[L(\phi)]}}. \tag{24}$$

## II. SEPARATION OF MOMENTS BY LO FIELD STRENGTH

The separation of moments using various LO field strengths $E_L$ works as follows. Consider $J$ pairs $(\tilde{E}_{L,j}, C(\tilde{E}_{L,j}; \phi))$, $j = 1, \ldots, J$, measured for a given phase $\phi$, where $\tilde{E}_{L,j} = E_{L,j}/E_{L,\text{ref}}$ is a rescaled LO field strength. One may choose, e.g., $E_{L,\text{ref}} = E_{L,k}$ with $k = 1, \ldots, J$. Since one has to measure anyway with blocked LO, one may include $\tilde{E}_{L,1} = 0$. Recalling Eq. (3) together with Eqs. (4)–(6) of the Letter, one has to apply a quadratic fit of the form

$$C(\tilde{E}_L; \phi) = C_0(\tilde{E}_L; \phi) + C_1(\tilde{E}_L; \phi) + C_2(\tilde{E}_L; \phi), \tag{25}$$

where $C_k(\tilde{E}_L; \phi)$ is proportional to $\tilde{E}_L^k$, i.e.,

$$C_k(\tilde{E}_L; \phi) = D_k(\phi)\tilde{E}_L^k. \tag{26}$$

The solution of the fitting problem (25) is given by

$$\boldsymbol{D} = (V^T V)^{-1} V^T \boldsymbol{C} \tag{27}$$

(see e.g. [1, 2]) with the Vandermonde matrix

$$V = \begin{pmatrix} 1 & \tilde{E}_{L,1} & \tilde{E}_{L,1}^2 \\ 1 & \tilde{E}_{L,2} & \tilde{E}_{L,2}^2 \\ \vdots & \vdots & \vdots \\ 1 & \tilde{E}_{L,J} & \tilde{E}_{L,J}^2 \end{pmatrix}, \tag{28}$$

$\boldsymbol{D} = (D_0(\phi), D_1(\phi), D_2(\phi))^T$, and $\boldsymbol{C} = (C(E_{L,1}; \phi), \cdots, C(E_{L,J}; \phi))$. It is necessary to measure at least for three ($J = 3$) different LO intensities including the blocked LO, however, more regression points ensure a more reliable separation of moments. The statistical error can be determined by

$$u_{D_k(\phi)} = \sqrt{\sum_{j=1}^{J} [((V^T V)^{-1} V^T)_{kj} (u_{\boldsymbol{C}})_j]^2}, \tag{29}$$

with $i = 0, \ldots, 2$. Although this way works in principle if the measured data points sufficiently coincide with the quadratic fit, it is very sensitive to systematic errors. That is why we apply a more robust analysis by using data from a second phase $\phi + \pi$ to infer the $D_k(\phi)$. In particular, the relations

$$D_0(\phi) = D_0(\phi + \pi) = C_{\text{block}} \tag{30}$$
$$D_1(\phi) = -D_1(\phi + \pi) \tag{31}$$
$$D_2(\phi) = D_2(\phi + \pi), \tag{32}$$

are always fulfilled. Defining

$$y(\tilde{E}_L; \phi) = \frac{C(\tilde{E}_L; \phi) - C_{\text{block}}}{\tilde{E}_L}, \tag{33}$$

yields the two linear equations

$$y(\tilde{E}_L; \phi) = D_2(\phi)\tilde{E}_L + D_1(\phi) \tag{34}$$
$$y(\tilde{E}_L; \phi + \pi) = D_2(\phi)\tilde{E}_L - D_1(\phi). \tag{35}$$

Through a coupled linear regression one obtains $D_1(\phi)$ and $D_2(\phi)$. In particular, following the method of least squares, the function

$$\mathcal{E} = \sum_{j=1}^{J} [y_j(\phi) - D_2(\phi)\tilde{E}_{L,j} - D_1(\phi)]^2 \tag{36}$$

$$+ \sum_{j=1}^{J} [y_j(\phi + \pi) - D_2(\phi)\tilde{E}_{L,j} + D_1(\phi)]^2 \tag{37}$$

has to be minimized. Minimization with respect to $D_1(\phi)$ and $D_2(\phi)$ yields

$$D_1(\phi) = \frac{1}{2J} \sum_{j=1}^{J} [y_j(\phi) - y_j(\phi+\pi)] \tag{38}$$

$$D_2(\phi) = \frac{\sum_{j=1}^{J} [y_j(\phi) + y_j(\phi+\pi)] \tilde{E}_{L,j}}{2 \sum_{j=1}^{J} \tilde{E}_{L,j}^2} \tag{39}$$

with the statistical uncertainties

$$u_{D_1(\phi)} = \frac{1}{2J} \sqrt{\sum_{j=1}^{J} \left[ u_{y_j(\phi)}^2 + u_{y_j(\phi+\pi)}^2 \right]} \tag{40}$$

$$u_{D_2(\phi)} = \frac{1}{2 \sum_{j=1}^{J} \tilde{E}_{L,j}^2} \sqrt{\sum_{j=1}^{J} \tilde{E}_{L,j}^2 \left[ u_{y_j(\phi)}^2 + u_{y_j(\phi+\pi)}^2 \right]}. \tag{41}$$

Here, the uncertainty $u_{y_j(\phi)}$ is composed of the statistical error

$$u_{y_j(\phi)} = \frac{1}{\tilde{E}_{L,j}} \sqrt{u_{C(E_{L,j};\phi)}^2 + u_{C_{\text{block}}}^2} \tag{42}$$

and an error due to a possible drift of the signal state, which is estimated by the deviation of $y_j$ from the quadratic fit. The error of the zeroth-order term is due to Eq. (30) given by

$$u_{D_0(\phi)} = u_{C_{\text{block}}}. \tag{43}$$

Similar to Eq. (11), for a proper application of our nonclassicality condition the quantities $D_k$ have to be converted to

$$\tilde{D}_k(\phi) = \frac{D_k(\phi)}{\mathcal{T}_k} \quad k = 0,1,2 \tag{44}$$

with the statistical uncertainty

$$u_{\tilde{D}_k(\phi)} = \frac{1}{|\mathcal{T}_k|} \sqrt{[\tilde{D}_k(\phi)]^2 u_{\mathcal{T}_k}^2 + u_{D_k(\phi)}^2} \tag{45}$$

and $u_{\mathcal{T}_k}$ determined according to Eqs. (12)–(14). A negative determinant of the matrix

$$\tilde{L}(\phi) = \begin{pmatrix} \tilde{D}_0(\phi) & \tilde{D}_1(\phi) \\ \tilde{D}_1(\phi) & \tilde{D}_2(\phi) \end{pmatrix} \tag{46}$$

indicates nonclassicality. Its error is obtained as

$$u_{\det[\tilde{L}(\phi)]} = \sqrt{[\tilde{D}_2(\phi)]^2 u_{\tilde{D}_0(\phi)}^2 + [\tilde{D}_0(\phi)]^2 u_{\tilde{D}_2(\phi)}^2 + 4[\tilde{D}_1(\phi)]^2 u_{\tilde{D}_1(\phi)}^2} \tag{47}$$

and the signed statistical significance is given by

$$\Sigma = \frac{\det[\tilde{L}(\phi)]}{u_{\det[\tilde{L}(\phi)]}}. \tag{48}$$

---


\end{widetext}

\end{document}